\documentclass[useAMS,usenatbib]{mn2e}
\usepackage{epsf}
\usepackage{psfig}

\title[The first spectroscopically confirmed Mira star in M\,33]
{The first confirmed Mira star in M\,33 \\
}
\author[E. A. Barsukova et al.]{E. A. Barsukova$^{1}$\thanks{E-mail:
bars@sao.ru (EAB)}, V. P. Goranskij$^{2}$,
K. Hornoch$^{3}$, S. Fabrika$^{1}$, W. Pietsch$^{4}$,
\newauthor O. Sholukhova$^{1}$, A. F. Valeev$^{1}$\\
$^{1}$Special Astrophysical Observatory of the Russian Academy of Sciences,
Nizhnij Arkhyz, Karachai-Cherkesia, 369167, Russia\\
$^{2}$Sternberg Astronomical Institute of the Moscow University,
Universitetskij prospect, 13, Moscow, 119992, Russia\\
$^{3}$Astronomical Institute, Academy of Sciences, CZ-251 65 Ond\u{r}ejov, Czech Republic\\
$^{4}$Max-Planck-Institut f\"ur extraterrestrische Physik, 85748 Garching, Germany}

\begin{document}

\date{Accepted 2010 December 22. Received 2010 September 14; in original form ...}

\pagerange{\pageref{firstpage}--\pageref{lastpage}} \pubyear{2010}
\maketitle

\label{firstpage}

\begin{abstract}
We present photometry and moderate-resolution spectroscopy of the luminous red
variable [HBS2006] 40671 originally detected as a possible nova in the
galaxy M\,33.
We found that the star is
a pulsating Mira-type variable with a long period
of 665 days and an amplitude exceeding 7 mag in $R$ band.
[HBS2006] 40671 is the first confirmed Mira-type star in M\,33.
It is one of the most luminous Mira-type variables. In the
$K$ band its mean absolute magnitude is M$_K=-$9.5, its bolometric magnitude
measured in the maximum light 
is also extreme, M$_{bol}=-$7.4. The spectral type of the star in the maximum
is M2e $-$ M3e. The heliocentric radial velocity of the star is $-475$~km/s.
There is a big negative excess ($-$210~km/s) in radial velocity of [HBS2006] 40671
relative to the average radial
velocity of stars in its neighborhood pointing at an exceptional
peculiar motion of the star.
All the extreme properties of the new Mira star make it
important for further studies.
\end{abstract}


\section{Introduction}

The General Catalogue of Variable Stars (GCVS) \citep{k1} gives a
determination of Mira (Omicron) Ceti-type variables:
``These are long-period variable giants with characteristic late-type
emission spectra (Me, Ce, Se) and
variation amplitudes from 2.5 to 11 magnitudes in $V$. Their periodicity is
       well pronounced, and the periods lie in the range between 80 and
       1000 days. Infrared amplitudes are usually less than those in the
       visible and may be $<$ 2.5 mag. For example, in the {\it K} band they
       usually do not exceed 0.9 mag.''
Due to the large amplitude of variability
exceeding 2.5 mag in the {\it V} band, Mira variables in nearby galaxies may
be confused with optical novae. The confusion can be solved, if subsequent
spectral
or photometric investigations reveal a Mira-type star with a cool M-type
spectrum and a long periodic variability.
Recently, peculiar red novae were introduced as
a new class of astrophysical objects \citep[e.g.,][]{g1}.
It is represented by stars such as V838 Mon,
V4332 Sgr and V1006/7 in M\,31. Nova Sgr 1943 (V1148 Sgr) may  possibly also
belong to this class.
Their remnants may be cool luminous L- or M-type supergiants,  which
in the course of discovery may be confused with both Mira-type stars
and classical novae.
Nova Sgr 1943 reached a maximum photometric brightness of 8.0 mag, it was described
by \citet{m1}
as a late K-type star with TiO bands in the spectrum.
It was found that the star is not of Mira-type, and later the star was lost.

In a search for novae in nearby galaxies one may run into similar problems.
A cool supergiant may be misidentified as a
nova when an unfiltered CCD image is compared with $B$ or $V$ images.
The M\,31 Nova candidate 2008-09b \citep{b1}
turned out to be a star with M9-type spectrum. It may be a red nova or even
a red supergiant showing no variability at all.
There are two additional nova candidates in M\,31 that turned out to be
luminous Mira-type stars:
M31N 1995-11e \citep{s1},
M31N 2007-11g \citep{s2}.
The Mira-type star in IC\,1613 described by \citet{k3}
has been discovered as a nova as well \citep{k2}.
In this paper we study a Mira-type star in the galaxy M\,33, which has been
discovered as a nova candidate.

\begin{figure}
\centerline{\hbox{\psfig{figure=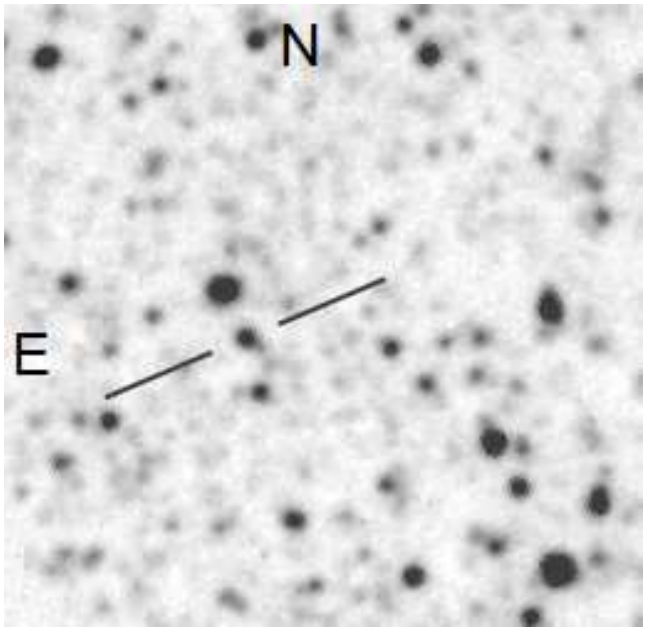,angle=0,clip=,width=8.5cm}}}
\caption{Finding chart of [HBS2006] 40671 made from the $R$ band Subaru
telescope frame taken on 2002 November 4. The new Mira-type star is
marked by two lines.
Size of the image is $25\arcsec \times 25\arcsec$.
}
\end{figure}

An outburst of a source with coordinates
$\rmn{RA} = 01^{\rmn{h}} 34^{\rmn{m}} 27\fs13$,
$\rmn{Dec} = 30\degr 58\arcmin 42\farcs 7$ (J2000) in M\,33  was detected
by \citet{n1}.
They reported it as a possible nova (with 18.6 mag) detected in
five frames taken around 2009 August 17.807 UT exposed by 60 sec using the
0.40-m f/9.8 reflector and unfiltered CCD of the Miyaki Argenteus Observatory,
Japan, reaching a limiting magnitude of 20.2.
Nothing was visible at this location on their previous frames taken on
2008 December 25.559 UT and 2009 August 07.737 UT with limiting magnitudes of 
20.0 and 19.4, respectively. The source was confirmed by these
authors on 2009 August 18.696.
But later they withdraw the interpretation as possible nova as it 
did not change brightness.

The source  was identified by us with a star number 40671 in
the photometric survey
of variable stars in M\,33 carried out by \citet{h1}. We will name it hereafter
[HSB2006] 40671. The survey observations were carried out between 2003 August
and 2005 January.
The authors classify the star as a Long Period Variable (LPV).
During their observations (filled circles in Fig. 3), the star was
red with $R - I \approx 2.7$ and $\approx 4.1$ mag at maximum and minimum
brightness, respectively.
The star became extremely red during the registered minimum ($R \ga 25$ mag).

\section{Observations}

We have carried out spectral observations
with the Russian 6-m BTA telescope using the SCORPIO spectrograph \citep{a1}.
The spectrum has been taken on 2009 October 9.874 UT in a range \hbox{4300 -
7880~\AA} with spectral resolution of 13~\AA. The spectroscopy was
followed by {\it BVR$_c$} images on the same date. We found the star at
a level of 18.8 mag  in the {\it R$_c$} band. One more photometric
{\it BVR$_c$} observation has been taken with the 1-m Zeiss telescope
of the Special Astrophysical Observatory (RAS) on 2009 October 20.9 UT.

We have searched for additional images of this star in Internet archives
of different observatories. Dozens of images were found, most of them come
from medium-size telescopes equipped with wide field mosaic CCD cameras.
The star position is well placed on the frames of the Calar Alto Observatory taken
in 1989 and 1998 with the 80/120-cm Schmidt telescope, on the frames taken in 2001
and 2002 in $BVRI$ bands with the 8.2-m SUBARU telescope of the National
Astronomical Observatory of Japan. One $V$ band observation from the
4.2-m William Herschel Telescope, and three SDSS $r'$ band observations
from the 2.54-m Isaac Newton Telescope were used as well, both telescopes
belong to the Isaac Newton Group of Telescopes at the Roque
de Los Muchachos Observatory at La Palma, Spain. We also apply one $I$ band
measurement obtained from the 4.0-m Mayall telescope
image taken in 2001 and published by \citet{m3}.

\begin{figure}
\centerline{\hbox{\psfig{figure=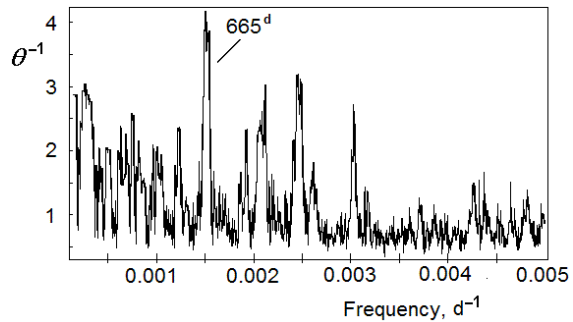,angle=0,clip=,width=7.5cm}}}
\caption{Periodogram of [HBS2006] 40671 calculated in the period
range between 200 and 7000 days using the \citet{l1}
phase dispersion minimization method.
$\theta$ is a normalized dispersion parameter.
}
\end{figure}

\begin{figure}
\centerline{\hbox{\psfig{figure=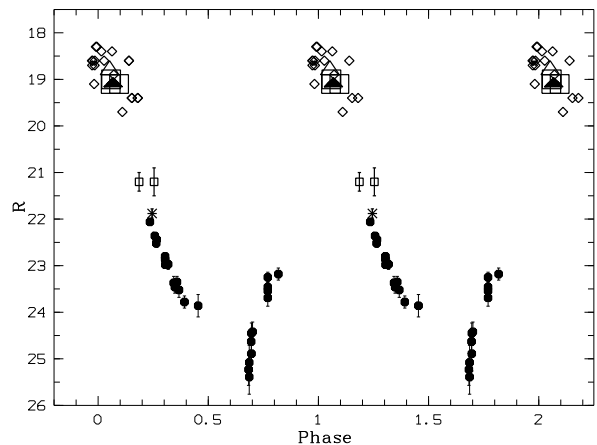,angle=-90,clip=,width=9.cm}}}
\caption{Light curve of [HBS2006] 40671 folded modulo
the 665-day period in R band. Filled circles are the
observations by \citet{h1}. Big open squares indicate observations from
Calar Alto,
the asterisk from the Subaru telescope,
open triangle from SAO BTA, filled triangle from SAO 1-m telescope,
small open squares are Sloan $r'$ magnitudes corrected by 0.6 mag.
Open rhombs are
unfiltered observations taken by \citep{n1}
(error bars $\sim 0.2$ mag are not shown for clarity).
At $R \ga 25$ mag the accuracy of the estimates is not better than 0.4 mag.
}
\end{figure}

\begin{table*}
\centering
\begin{minipage}{110mm}
\caption{Photometric Observations of [HBS2006] 40671.}
\begin{tabular}{@{}llccl@{}}
\hline
Date UT &      JD 24... & Magnitude  &     Band\footnote{For the plot of the
light curve (Fig. 3), we subtracted 0.6 mag
from Sloan {\it r~\arcmin} magnitudes to transform them into $R$ magnitudes
(using color transformations from \citet{j2} and assuming an average color index
$V-R = 1.2$ mag for the Mira-type star).}
&        Telescope\\
\hline
1989-10-03.11& 47802.61&   19.1$\pm$0.1  &   {\it R}         & 80/120 cm Calar Alto\\
1989-10-04.99& 47804.49&   19.0$\pm$0.1  &   {\it R}         & 80/120 cm Calar Alto\\
1998-10-29.03& 47828.53&   19.1          &  unfiltered & 80/120 cm Calar Alto\\
2001-09-18.27& 52170.77&   20.66$\pm$0.09&   {\it I}         & 4.0 m Mayall       \\
2001-11-20.22& 52233.72&   19.2$\pm$0.1  &   {\it I}         & 8.2 m SUBARU       \\
2001-11-20.39& 52233.89&   22.6$\pm$0.2  &   {\it V}         & 8.2 m SUBARU       \\
2001-11-20.46& 52233.96& $>$23.6         &   {\it B}         & 8.2 m SUBARU       \\
2002-09-26.06& 52543.56&   21.8$\pm$0.2  &   {\it r~\arcmin} & 2.54 m Sloan INT   \\
2002-11-04.23& 52582.73&   21.88$\pm$0.1 &   {\it R}         & 8.2 m SUBARU       \\
2002-11-10.09& 52588.59&   21.8$\pm$0.3  &   {\it r~\arcmin} & 2.54 m Sloan INT   \\
2004-09-10.08& 53258.58& $>$23.1         &   {\it V}         & 4.2 m WHT          \\
2008-10-09.12& 54748.62& $>$24.4         &   {\it r~\arcmin} &  2.54 m Sloan INT   \\
2009-08-17.81& 55061.31&   18.6          &  unfiltered & 0.40 m f/9.8 \\
2009-08-18.70& 55062.20&   18.7          &  unfiltered & 0.40 m f/9.8 \\
2009-08-19.75& 55063.25&   18.6          &  unfiltered & 0.40 m f/9.8 \\
2009-08-24.69& 55068.19&   19.1          &  unfiltered & 0.40 m f/9.8 \\
2009-08-25.71& 55069.21&   18.7          &  unfiltered & 0.40 m f/9.8 \\
2009-08-26.79& 55070.29&   18.6          &  unfiltered & 0.40 m f/9.8 \\
2009-08-29.77& 55073.27&   18.3          &  unfiltered & 0.40 m f/9.8 \\
2009-09-01.80& 55076.30&   18.3          &  unfiltered & 0.40 m f/9.8 \\
2009-09-15.69& 55090.19&   18.4          &  unfiltered & 0.40 m f/9.8 \\
2009-09-23.76& 55098.26&   18.6          &  unfiltered & 0.40 m f/9.8 \\
2009-10-09.8 & 55114.35&   21.8$\pm$0.1  &   {\it B}         & 6.0 m BTA SAO RAS     \\
2009-10-09.8 & 55114.36&   20.02$\pm$0.05&   {\it V}         & 6.0 m BTA SAO RAS     \\
2009-10-09.8 & 55114.36&   18.80$\pm$0.02&   {\it R}         & 6.0 m BTA SAO RAS     \\
2009-10-17.64& 55122.14&   18.4          &  unfiltered & 0.40 m f/9.8 \\
2009-10-20.91& 55125.41&   21.8$\pm$0.15 &   {\it B}         & 1.0 m SAO RAS         \\
2009-10-20.91& 55125.41&   20.32$\pm$0.07&   {\it V}         & 1.0 m SAO RAS         \\
2009-10-20.92& 55125.42&   19.06$\pm$0.04&   {\it R}         & 1.0 m SAO RAS         \\
2009-10-22.74& 55127.24&   18.9          &  unfiltered & 0.40 m f/9.8 \\
2009-11-23.53& 55153.03&   19.7          &  unfiltered & 0.40 m f/9.8 \\
2009-12-07.53& 55173.03&   18.6          &  unfiltered & 0.40 m f/9.8 \\
2009-12-16.56& 55182.06&   19.4          &  unfiltered & 0.40 m f/9.8 \\
2010-01-03.52& 55200.02&   19.4          &  unfiltered & 0.40 m f/9.8 \\
\hline
\end{tabular}
\end{minipage}
\end{table*}

Archival images from the 8.2-m SUBARU telescope, the 4.2-m William Herschel
Telescope, and the 2.54-m Isaac Newton Telescope were downloaded as raw FITS
files and then processed in the same way as our own 6-m BTA and \hbox{1-m} Zeiss
telescope images. Standard reduction procedures for raw CCD images were applied
(bias and dark-frame subtract and flat-field correction) using the
SIMS\footnote{\tt http://ccd.mii.cz/}
program. Reduced images of the same
series were co-added to improve the S/N ratio and then used for photometry.
We used GAIA\footnote{\tt http://www.starlink.rl.ac.uk/gaia}
to perform
``Optimal photometry'' (based on fitting of PSF profiles).
$B, V, R$, and $I$ magnitudes for comparison stars were taken from the Local
Group Galaxy Survey (LGGS) catalog of stars in M\,33
\citep{m3}.

In the case of SDSS $r'$ band images, we computed $r'$ magnitudes for
comparison stars from $BVRI$ magnitudes taken from 
\citet{m3} using empirical color transformations between
the SDSS $u'g'r'i'z'$ system and Johnson-Cousins $UBVRI$ system published by
\citet{j2}.

\citet{n1} continued their unfiltered observations at the Miyaki Argenteus
Observatory, Japan,
from 2009 August 18 to 2010 January 3 and kindly provided us with
16 new images taken in this period. All the photometric measurements
of [HBS2006] 40671 \citep[collected in addition to those of][]{h1} are
presented in Table 1.

We created a finding chart for the star from high-quality $R$ band image taken
with the Subaru telescope on 2002 November 4 (Fig. 1).
A relatively bright visual companion is located 2.1 arcsec
north-east of the variable.
\citet{m3}
cataloged the position of this star as $\rmn{RA} = 01^{\rmn{h}} 34^{\rm
n{m}} 27\fs19$,
$\rmn{Dec} = 30\degr 58\arcmin 44\farcs 4$ (J2000) and give $B$, $V$, $R$, and $I$
band magnitudes of 21.40, 21.11, 20.94, and 20.71, respectively.
The observations of the LGGS \citep{m3} were obtained from
October 2000 to September 2001.
The variable star is not included in the catalog, as it is only detected
on the $I$ band image taken during the survey on 2001 September 18
(criterion for inclusion in catalog is detection in the three bands $B,V,R$).

\begin{figure*}
\centerline{\hbox{\psfig{figure=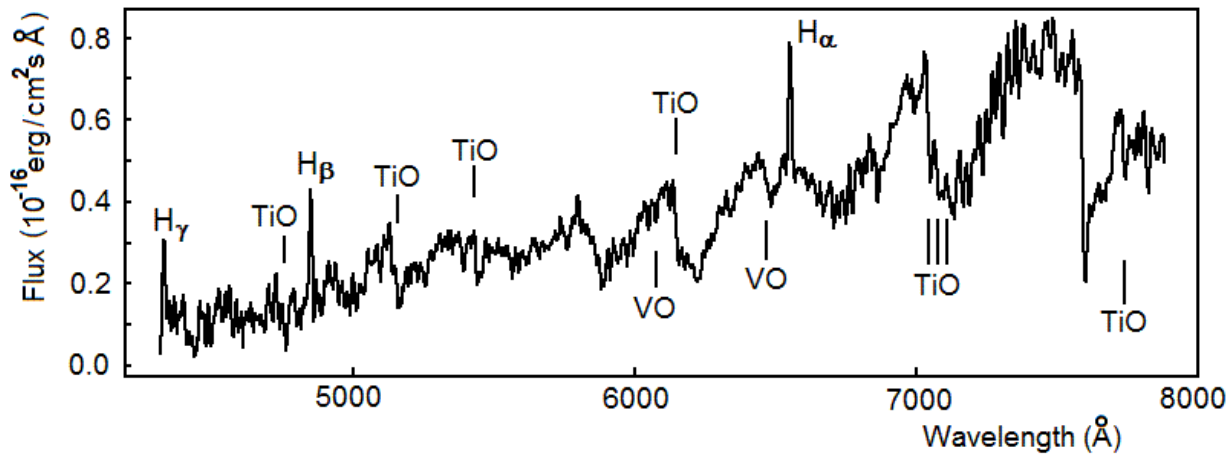,angle=0,clip=,width=16.5cm}}}
\caption{The spectrum of [HBS2006] 40671 taken with BTA on
2009 October 9 near the maximum light.
}
\end{figure*}

\section{The period search and the light curve}

Most of the data in Table 1 relate to the light curve maximum.
Both ascending and descending branches of the light curve are not fully covered.
The observations during these phases come mostly from the survey
by \citet{h1}. The survey presents observations
when the star had a brightness between 22 and 25 mag in the $R$ band.
The accuracy of these observations \citep{h1} varies from 0.04 mag
at $R$ = 22 mag to 0.37 mag at $R$ = 25 mag.

To search for periodicities we used a method proposed by \citet{l1}
which is based on the phase
dispersion minimization. In this search we used only the observations
in {\it R} band because they represent the best sampling among the total data set.

The periodogram was calculated in the period range between
200 and 7000 days (Fig. 2). The output parameter there is
the inverse dispersion value $1/\theta$ described by \citet{l1},
maxima in the periodogram correspond to dispersion minima.
There are essentially no peaks with period values less than 200 days.
This range of periods is not shown in the figure.
A period of 665 days shows the highest amplitude.
There are two lower amplitude peaks at 3500 and 406 days (Fig. 2).
However, these periods give much worse light curves.
Longer periods contradict with the fast changes of brightness
on ascending and descending branches of the light curve \citep{h1}.
The phased light curve in the $R$ band is presented in Fig. 3. It was
calculated with the following light elements
\begin{equation}
Max = 2455080 + 665 \ast E.
\end{equation}
\label{efemeries}

\noindent
Figure 3 shows that we do not resolve and possibly not even cover
the minimum of the light curve. The
star in minimum brightness is not brighter than $R$ = 25.4 mag.
The total amplitude
of the variability is bigger than 7 mag in the $R$ band.
Also the ascending branch is not well covered.
Therefore,
the epoch of the maximum can only be constrained to $\pm5$ days.
However, it is obvious that the light curve of this Mira variable is
asymmetric, the ascending branch is steeper than the
descending one.

We tried to refine the maximum in the light curve using archival observations
of the infrared satellite Spitzer \citep{m2}.
In five observations on 2004 January 09, July 22,
August 16, and 2005 January 21 and August 25, the star varied in the range of
14.24 -- 14.97 mag,
14.16 -- 14.64 mag, and 13.66 - 14.04 mag in the 3.6, 4.5, and 8 $\,\umu$m bands,
respectively. A maximum is found during July/August 2004 corresponding to 
phases in the interval
0.18--0.22
according to Equation 1.
Therefore the infrared maximum seems not to be in phase with the optical
maximum exactly, but it is very close to it (see Fig. 3).

\section{Spectrum}

The calibrated spectrum of [HBS2006] 40671 in the 4300 - 7880 \AA\ spectral
region is shown in Fig. 4.
Molecular TiO bands dominate in the spectrum.
We classify the star as an oxygen-rich (M-type or O Mira)
star \citep{s3,j1}.
Strong narrow emission lines of hydrogen are present as well; this is typical for
Mira stars during maximum light. Using the spectrum we estimate a
contribution of the TiO bands into the star brightness in $B$ and $V$ bands.
They are $\delta(B) = -$0.28 mag and $\delta(V) = -$0.16 mag.
These values agree with the spectral type M1 -- M2 \citep{s3}.

We have also determined the spectral type of the star using [TiO]$_1$ and [TiO]$_2$
indices introduced by \citet{o1} and
measured in our calibrated spectrum.
The spectral indices were derived from fluxes
measured in 30~\AA\ bandpasses centered at three wavelengths each for
[TiO]$_1$ (6125, 6180 and 6370 \AA) and for [TiO]$_2$ (7025, 7100 and 7400 \AA)
as [TiO]$_1 = +0.52$ and
[TiO]$_2 = +0.55$, respectively. These index values correspond to spectral class M3.
Such a spectrum is natural for Mira variables in maximum light, while in
the minimum they have usually later spectra.

The H$\alpha$ emission line of [HBS2006] 40671 is very narrow and not
resolved in our spectrum. Its equivalent width is EW = 9 \AA.
For H$\alpha$ we measure a heliocentric radial velocity of $-440 \pm15$ km/s.
That this high radial velocity is not caused by calibration problems was
verified by measuring the radial velocity of the [OI] $\lambda$6300 sky line
which was determined as \hbox{0 km/s} as expected.
Also, at a distance of $\sim 1.\arcmin6$ from [HBS2006] 40671, by chance
the high ionization HII region Z\,378 \citep{curt}
was captured in the slit.
The heliocentric radial velocity of its H$\alpha$ line is $-280 \pm10$ km/s.
This velocity is in good agreement with that of the HI radio emission
at the position \citep [$-265 \pm5$ km/s,][]{r1}.
The difference of  $1.6'$ between the HII region and the star's
location in the galaxy body may produce a radial velocity shift
of 5 -- 10 km/s.
Therefore we find a strong negative excess in the H$\alpha$
radial velocity of [HBS2006] 40671 of $-175$ km/s over the galaxy radial velocity.

To check whether this negative excess of the H$\alpha$ radial velocity
also is reflected in other features of the spectrum of [HBS2006] 40671,
we cross-correlated it with spectra of known standard stars.
We used two red variables, V934 Her (HD 154791, 4U 1700+24, 
M2 III) and KK Per (HD 13136, M2 Iab-Ib). The cross-correlation region was
5850 -- 6700 \AA \, excluding H$_\alpha$ emission and interstellar NaI lines.
The spectrum of V934 Her was taken with the SAO 1-m Zeiss telescope 
on 2006 April 6.95 UT. According to \citet{g2} it is an X-ray binary with
a period of P = 404 $\pm$3~days.
The radial velocity amplitude of the primary is only K = 0.75 $\pm$0.12~km/s.
The systemic velocity of V934 Her is  $-48.7 \pm$0.1~km/s.
The cross-correlation resulted in a radial velocity of $-474$~km/s for the Mira star.
We obtained almost the same value ($-477$~km/s) using the second standard star
KK Per \citep[its radial velocity is $-39.4 \pm$0.4~km/s,][]{m0}.

We conclude that the star has a big negative peculiar radial velocity compared to
the body of the galaxy (HI data) of $-210 \pm 10$~km/s. Its H$_\alpha$ emission
is redshifted relative to the star by $+35 \pm 10$~km/s. Such a shift
might be explained by stellar pulsations. However, it could also be
caused by the superposition of the emission line on an
absorption line that is not resolved in our spectrum. The huge peculiar velocity
can not be explained by the star binarity. To get an orbital velocity of
$\sim 200$~km/s in a few solar mass binary
the orbital period must be less than 10 days, and the orbital separation less than 
0.2~a.u. This is impossible due to the huge sizes of Miras.

We tried to investigate if [HBS2006] 40671 is positioned  in front of or behind
the disk of M\,33 which would indicate motion away from or to the disk,
respectively. However, the signal to noise of the spectrum in the range of
expected diffuse interstellar bands \citep[e.g.][]{Cordiner}
was not sufficient.

\section{Discussion and conclusion}

According to our data from the 6-m and 1-m telescopes, the observed
photometric colours of [HBS2006] 40671 during maximum are $B-V = 1.6$ and
$V-R = 1.2$ mag.
Taking into account the distance modulus of M\,33
$(m-M)_0 = 24.92 \pm0.12$ mag \citep[the distance $964 \pm54$ kpc,][]{b2}
and the Galactic extinction value $A_V = 0.15$ mag ($A_K = 0.02$ mag) in
that direction \citep{s4}
we find an absolute magnitude of the star near maximum brightness
(V$ = 20.02 \pm 0.05$ mag at the pulsation phase 0.05)
of M$_V = -5.05 \pm 0.15$. This is consistent with luminosity classes Ib--Iab.
For an M3-type star the bolometric correction amounts to $-$2.34 mag
\citep{s5}. This leads to
a bolometric magnitude  in the brightness maximum of M$_{bol} \approx -7.4$.
Unfortunately, we have not enough data to estimate
$V$ band pulsation magnitude or mean magnitude.

2MASS data \citep{s6} show [HBS2006] 40671 on 1997 December 5, in the pulsation
phase of 0.54, at a brightness of 17.30 $\pm 0.23$,
15.89 $\pm 0.16$ and 15.45 $\pm 0.15$ mag in $J,H$ and $K$, respectively.
There are more observations of the star in the $JHK$ bands taken from 2005
September 29 to December 16 (pulsation phases 0.84--0.96)
with the 3.8-m UKIRT telescope equipped with WFCAM \citep{c1}.
The brightness of the star during this period was
16.86, 16.24 and 15.66 mag in the $JHK$ bands, respectively.
These data confirm the typical Mira behaviour, i.e. the decrease of pulsation
amplitude with increasing wavelength.
However we have not enough data to study the infrared light curve in detail.
\citet{Kanbur1997} found that the $K$ band amplitudes of O Mira variables
is smaller for shorter periods.
From the relations in that paper,
the O Mira variables with periods longer than 420 days should show
total $K$ amplitudes smaller than 0.4 mag.
We use the 2MASS data for a luminosity estimate. We find an
absolute magnitude of the [HBS2006] 40671 as M$_K \sim -9.5$ with a possible
error of \hbox{$\pm$ 0.15}. 
Applying a bolometric correction of $+3.3$ mag
determined by \citet{b0} to this $K$ band magnitude we find M$_{bol} = -6.2$.

\citet{f1} present period - luminosity (PL) relations for LMC Mira variables,
where they found that the O Mira variables with periods longer
than $\sim 420$~days are overluminous \citep[see also][]{h2}. Fitting the
overluminous branch of the long-period Mira variables \citep[without the star C2 whose
status is less certain,][]{f1} we find for our star with the 665-day
period that the expected mean luminosity is M$_K \sim -9.47$ mag. This agrees
well with our estimate from the 2MASS data.
Using PL relations for bolometric luminosities for the overluminous
long-period Mira variables both from \citet{f1} and \citet{h2} we find very
similar results, the mean expected luminosity for the 665-day star
is M$_{bol} \approx -6.5$ mag. 
This value is close to M$_{bol} = -6.2$,
found above from the 2MASS data taken in the minimum. 
On the basis of our spectral classification,
we determined the bolometric luminosity of $\approx -7.4$ mag is
in the brightness maximum. The bolometric luminosities of O-type Mira variables
are notably changing during the pulsation cycle. Using the relations found
by \citet{Kanbur1997} one may expect that in O Mira variables with periods longer
than 420 days the difference between maximal and mean bolometric luminosity
is larger than 0.5 mag. We conclude that luminosities of [HBS2006] 40671
found by us are in line with the
statistical relations, but the star's parameters put it
in the extreme tail of these relations.

In the presence of the huge amplitude of [HBS2006] 40671 in the $R$ band,
not less than 7 mag, one may expect an even larger
amplitude in the $V$ band \citep[e.g.][]{Barthes1996}.
In the long-period Mira variables the amplitude dispersion is very high \citep{m4,h2}.
A visual amplitude in a Mira star with a period P $\sim$ 665~days may be as
large
as 8.3 mag \citep{m4}. Using relations presented by \citet{h2} we estimate
a mass of this star as M $\sim$ 4 solar masses. Based on its luminosity and
the pulsation period, [HBS2006] 40671 is similar to the Mira-type
variable discovered by \citet{k3} in the IC\,1613 galaxy (M$_K = -9.62$ mag and $
P = 640.7$~day).
However the amplitude of the IC\,1613 Mira is only $2.5 - 3$ mag in $R$ band.
In the PL plane for M$_K$ luminosities for LMC the star [HBS2006] 40671 is
located in the same place as the IC\,1613 Mira \citep{k3}. Both stars are
located in the zone of the first-overtone pulsating Mira variables.

\citet{w1} found that long-period variables are grouped into two classes:
core-helium-burning supergiants which are brightest, and AGB stars which
are at least \hbox{1 mag} fainter at a given period. The supergiant LPVs
form a distinct PL relation and they have lower amplitudes. The
PL relation for six core-helium-burning supergiants in M\,33 was
presented by \citet{k4} in their Fig.~8. The $K$ magnitude for supergiants with
the 665 day period is expected to be 13.9. The 2MASS $K$ magnitude of [HBS2006] 40671 is
1.5 mag fainter, what suggests that this object is an AGB star. AGB stars in
M\,33 and M\,31 are discussed in papers by \citet{j0} and \citet{r2}.

\citet{w0} discussed a type of long-period variables which are
undergoing a hot bottom burning (HBB) in the base of their convective envelopes.
They lie above the PL
relation of AGB Mira-type variables. In LMC they are large amplitude variable
stars additionally showing a lithium and s-process enrichment. As an
example, she refers to 641-day Mira in IC 1613 studied by \citet{k3} that locates
above the PL relation. [HBS2006] 40671 also locates 0.35 mag above the PL relation
for AGB stars with periods over 400 days derived by \citet{h2} along with
three HBB stars studied by \citet{w3} in LMC. We did not find lithium
or other s-process elements in our low-resolution spectrum, but the luminosity
excess is evident as that in the IC 1613 Mira variable.

We conclude that the star [HBS2006] 40671 in M\,33 is O Mira-type variable with
extreme properties. Its pulsation amplitude is not less than 7 magnitudes
in $R$ band, the period of $\sim$ 665~days is one of the longest known for
Mira-type variables.
The mean absolute magnitude of the star in $K$ band is M$_K \sim -9.5$.
In maximum light
its bolometric magnitude is estimated as
M$_{bol} \approx -7.4$. It shows a spectrum of type M2e--M3e.

There is a strong negative excess of $-$210~km/s in the star velocity
relative to the radial velocity of the star's location projected to the
galactic disk. This is a peculiar motion of the star.
According to \citet{Feast} the velocity dispersion of Galactic O-Miras is in
the range 30 to 80~km/s. They belong to the galactic populations ranging from
the thin disk to the extended disk. In spite of the enhanced velocity dispersion
of the Galactic Mira-stars the peculiar velocity of $\sim -$200~km/s in the
[HBS2006] 40671 is rather big.

The radial velocity of the narrow H$\alpha$\ emission line is
$+35 \pm 10$~km/s relative to that of the star itself. Such a shift in the
H$\alpha$\ emission is typical for Mira stars. It can be explained by  
stellar pulsations and shocks in an expanding atmosphere.
All the properties make the new Mira star important for further studies.
[HBS2006] 40671 is the first spectroscopically
confirmed Mira star in M\,33.

\section*{Acknowledgments}

We are grateful to K. Nishiyama and F. Kabashima for providing us with
their unpublished
photometrical data.
We thank Joanna Mikolajewska for useful discussion.
Authors thank the Russian Foundation for Basic
Research for support
by grants No. 07-02-00630, 09-02-00163 and 10-02-00463,
Federal Programme ``Scientific and
educational cadre of innovating Russia 2009--2013'', No. 1244
and the grant ``Leading Scientific Schools
of Russia'' No. 5473.2010.2.
We gratefully acknowledge the support of the HDAP which was produced
at Landessternwarte
Heidelberg-Koenigstuhl under grant No. 00.071.2005 of the
Klaus-Tschira-Foundation.
This paper makes use of data obtained from the Isaac Newton Group
Archive which is maintained as part of the CASU Astronomical Data Centre
at the Institute of Astronomy, Cambridge. It is also based  on data
collected at Subaru Telescope and obtained from the SMOKA, which
is operated by the Astronomy Data Center, National Astronomical Observatory
of Japan.
This research has made use of the SIMBAD database, operated at CDS,
Strasbourg, France, and of NASA's Astrophysics Data System Bibliographic
Services.


\begin{thebibliography}{}
\bibitem[\protect\citeauthoryear{Afanasiev \& Moiseev}
{2005}]{a1}
Afanasiev V.~L., Moiseev A.~V., 2005, AstL, 31, 194

\bibitem[\protect\citeauthoryear{Barsukova et al.}%
{2008}]{b1}
Barsukova E., Fabrika S., Sholukhova O.,
Valeev A., Goranskij V., Pietsch W., Hornoch K., 2008, ATel, 1762

\bibitem[\protect\citeauthoryear{Barthes, Chenevez, \& Mattei}{1996}]
{Barthes1996}
Barthes D., Chenevez J., Mattei J.~A., 1996, AJ, 111, 2391 

\bibitem[\protect\citeauthoryear{Bessell \& Wood}{1984}]
{b0}
Bessell M.~S. \& Wood P.~R., 1984, PASP, 96, 247

\bibitem[\protect\citeauthoryear{Bonanos et al.}%
{2006}]{b2}
Bonanos A.~Z., et al., 2006, ApJ, 652, 313

\bibitem[\protect\citeauthoryear{Cioni et al.}
{2008}]{c1}
Cioni M.-R.~L., et al., 2008, A\&A, 487, 131

\bibitem[\protect\citeauthoryear{Cordiner et
al.}{2010}]{Cordiner}
Cordiner M.~A., Cox N.~L.~J., Evans C.~J.,
Trundle C., Smith K.~T., Sarre P.~J., Gordon K.~D., 2010, arXiv,
arXiv:1011.0797

\bibitem[\protect\citeauthoryear{Courtes et al.}
{1987}]{curt}
Courtes G., Petit H., Sivan J.-P., Dodonov S., Petit M., 1987, A\&A,
174, 28

\bibitem[\protect\citeauthoryear{Feast et al.}%
{1989}]{f1}
Feast M.~W., Glass I.~S., Whitelock P.~A., Catchpole R.~M., 1989, MNRAS,
241, 375

\bibitem[\protect\citeauthoryear{Feast}%
{2007}]{Feast}
Feast M., 2007, ASPC, 378, 479

\bibitem[\protect\citeauthoryear{Galloway et al.}%
{2002}]{g2}
Galloway D.~K., Sokoloski J.~K., Kenyon S.~J., 2002, ApJ, 580, 1065

\bibitem[\protect\citeauthoryear{Goranskij \& Barsukova}%
{2007}]{g1}
Goranskii V.~P., Barsukova E.~A., 2007, ARep, 51, 126

\bibitem[\protect\citeauthoryear{Hartman et al.}%
{2006}]{h1}
Hartman J.~D., Bersier D., Stanek K.~Z.,
Beaulieu J.-P., Kaluzny J., Marquette J.-B., Stetson P.~B.,
Schwarzenberg-Czerny A., 2006, MNRAS, 371, 1405

\bibitem[\protect\citeauthoryear{Hughes \& Wood}
{1990}]{h2}
Hughes S.~M.~G., Wood P.~R., 1990, AJ, 99, 784

\bibitem[\protect\citeauthoryear{Jaschek \& Jaschek}%
{1987}]{j1}
Jaschek C., Jaschek M., 1987, JBAA, 98, 47

\bibitem[\protect\citeauthoryear{Javadi et al.}%
{2010}]{j0}
Javadi A., van Loon J.~Th., Mirtorabi M.~T., 2010, arXiv, arXiv:1009.1822

\bibitem[\protect\citeauthoryear{Jordi et al.}%
{2006}]{j2}
{Jordi K., Grebel E.~ K., Ammon, K., 2006, A\&A, 460, 339}

\bibitem[\protect\citeauthoryear{Kanbur et al.}{1997}]{Kanbur1997}
Kanbur S.~M., Hendry M.~A., Clarke D., 1997, MNRAS, 289, 428

\bibitem[\protect\citeauthoryear{Kholopov et al.}%
{1985}]{k1}
Kholopov P.~N. et al., 1985, General Catalogue
of Variable Stars, 4th edition, Nauka Publishing House, Moscow [in Russian]

\bibitem[\protect\citeauthoryear{King et al.}%
{1999}]{k2}
King J.~Y., Modjaz M., Li W.~D., 1999, IAUC, 7287, 2

\bibitem[\protect\citeauthoryear{Kinman et al.}%
{1987}]{k4}
Kinman T.~D., Mould J.~R., Wood P.~R., 1987, AJ, 93, 833

\bibitem[\protect\citeauthoryear{Kurtev et al.}%
{2001}]{k3}
Kurtev R., Georgiev L., Borissova J., Li W.~D., Filippenko A.~V.,
Treffers R.~R., 2001, A\&A, 378, 449

\bibitem[\protect\citeauthoryear{Lafler \& Kinman}%
{1965}]{l1}
Lafler J., Kinman T.~D., 1965, ApJS, 11, 216

\bibitem[\protect\citeauthoryear{Mayall}
{1949}]{m1}
Mayall M.~W., 1949, AJ, 54, 191

\bibitem[\protect\citeauthoryear{McQuinn et al.}%
{2007}]{m2}
McQuinn K.~B.~W., et al., 2007, ApJ, 664, 850

\bibitem[\protect\citeauthoryear{Marrese et al.}%
{2003}]{m0}
Marrese P.~M., Boschi F., Munari U., 2003, A\&A, 406, 995

\bibitem[\protect\citeauthoryear{Massey et al.}%
{2006}]{m3}
Massey P., Olsen K.~A.~G., Hodge P.~W., Strong S.~B., Jacoby G.~H.,
Schlingman W., Smith R.~C., 2006, AJ, 131, 2478

\bibitem[\protect\citeauthoryear{Mattei et al.}
{1997}]{m4}
Mattei J.~A., Foster G., Hurwitz L.~A., Malatesta K.~H., Willson L.~A.,
Mennessier M.~O., 1997, ESASP, 402, 269

\bibitem[\protect\citeauthoryear{Nishiyama \& Kabashima}%
{2009}]{n1}
Nishiyama K., Kabashima F., private comm., 2009

\bibitem[\protect\citeauthoryear{O'Connell}
{1973}]{o1}
O'Connell R.~W., 1973, AJ, 78, 1074

\bibitem[\protect\citeauthoryear{Reakes \& Newton}
{1978}]{r1}
Reakes M.~L., Newton K., 1978, MNRAS, 185, 277

\bibitem[\protect\citeauthoryear{Rich et al.}
{1993}]{r2}
Rich R.~M., Mould J.~R., Graham J.~R., 1993, AJ, 106, 2252

\bibitem[\protect\citeauthoryear{Schlegel et al.}
{1998}]{s4}
Schlegel D.~J., Finkbeiner D.~P., Davis M., 1998, ApJ, 500, 525


\bibitem[\protect\citeauthoryear{Shafter et al.}
{2008a}]{s1}
Shafter A.~W., Ciardullo R., Bode M.~F.,
Darnley M.~J., Misselt K.~A., Nishiyama K., Kabashima F., 2008a, ATel, 1834

\bibitem[\protect\citeauthoryear{Shafter et al.}
{2008b}]{s2}
Shafter A.~W., et al., 2008b, ATel, 1851


\bibitem[\protect\citeauthoryear{Skrutskie et
al.}{2006}]{s6}
Skrutskie M.~F., et al., 2006, AJ, 131, 1163

\bibitem[\protect\citeauthoryear{Smak}
{1964}]{s3}
Smak J., 1964, ApJS, 9, 141

\bibitem[\protect\citeauthoryear{Smak}
{1966}]{s5}
Smak J., 1966, AcA, 16, 1

\bibitem[\protect\citeauthoryear{Whitelock et al.}
{2003}]{w3}
Whitelock P.~A., Feast M.~W., van Loon J.~Th., Zijlstra A.~A., 2003, MNRAS, 342, 86

\bibitem[\protect\citeauthoryear{Whitelock}{2010}]{w0}
Whitelock P.~A., 2010, in Variable Stars, the Galactic Halo and Galaxy Formation.
B.~V. Kukarkin Centennary Conference. Ed. C. Sterken, N. Samus, and L. Szabados.
Publ. by SAI and Moscow University. P.55

\bibitem[\protect\citeauthoryear{Wood et al.}{1983}]{w1}
Wood P.~R., Bessell M.~S., Fox M.~W., 1983, ApJ, 272, 99


\end{thebibliography}
\end{document}